\documentclass[aps,prl,showpacs]{revtex4}
\usepackage{amssymb}

\usepackage{amsmath}
\usepackage{graphicx}
\usepackage{bm}

\begin{document}

\title{A new class of quasi-exactly solvable potentials with position
dependent mass}
\date{\today}
\author{Ramazan Ko\c{c}}
\email{koc@gantep.edu.tr}
\affiliation{Department of Physics, Faculty of Engineering University of Gaziantep, 27310
Gaziantep, Turkey}
\author{Mehmet Koca}
\email{kocam@squ.edu.om}
\affiliation{Department of Physics, College of Science, Sultan Qaboos University, PO Box
36, Al-Khod 123, Muscat, Sultanate of Oman}
\author{Eser K\"{o}rc\"{u}k}
\email{korcuk@gantep.edu.tr}
\affiliation{Department of Physics, Faculty of Engineering University of Gaziantep, 27310
Gaziantep, Turkey}

\begin{abstract}
A new class of quasi exactly solvable potentials with a variable mass in the
Schr\"{o}dinger equation is presented. We have derived a general expression
for the potentials also including Natanzon confluent potentials. The general
solution of the Schr\"{o}dinger equation is determined and the eigenstates
are expressed in terms of the orthogonal polynomials.
\end{abstract}

\maketitle

In recent years, physical systems with a position dependent mass\cite{foul,
roy, dutra} and quasi exactly solvable(QES) potentials\cite{turb} have been
the focus of interest. The effective mass models have been used to describe
electronic properties of semiconductors, liquid crystals and various other
physical systems\cite{serra}. In this letter we suggest a method to obtain a
general solution of the Schr\"{o}dinger equation with a position dependent
mass.

We start with a general Hermitian effective mass Hamiltonian which is
proposed by von Roos\cite{roos}%
\begin{equation}
H=\frac{1}{4}\left( m^{\alpha }(x)\mathbf{p}m^{\beta }(x)\mathbf{p}m^{\gamma
}(x)+m^{\gamma }(x)\mathbf{p}m^{\beta }(x)\mathbf{p}m^{\alpha }(x)\right)
+V(x)  \label{eq:1}
\end{equation}%
with the constraint over the parameters: $\alpha +\beta +\gamma =-1$.
Depending on the choice of parameters the Hamiltonian(\ref{eq:1}) can be
expressed in different forms\cite{roy}. However, we shall keep the general
form of the Hamiltonian. Using the differential operator equivalence of
momentum operator $\mathbf{p=-}i\frac{d}{dx}$, it is easy to show that the
Hamiltonian (\ref{eq:1}) can be written as%
\begin{eqnarray}
-\frac{1}{2m(x)}\frac{d^{2}\psi (x)}{dx^{2}}+\frac{m^{\prime }(x)}{2m^{2}(x)}%
\frac{d\psi (x)}{dx}+(V(x)-E)\psi (x)+ &&  \notag \\
\left[ (1+\beta )m(x)m^{\prime \prime }(x)-2(\beta +1+\alpha (\alpha +\beta
+1)m^{\prime 2}(x)\right] \frac{\psi (x)}{4m^{3}(x)}=0 &&  \label{eq:2}
\end{eqnarray}%
where $E$ is the eigenvalue of the Hamiltonian (\ref{eq:1}). Our task is now
to obtain a general expression for the potential $V(x)$ such that the Schr%
\"{o}dinger equation can be solved quasi-exactly. Without loss of
generality, consider the following QES second order differential equation%
\cite{koc},%
\begin{equation}
z\frac{d^{2}\Re (z)}{dz^{2}}+\left( \ell +\frac{3}{2}+z(b-qz)\right) \frac{%
d\Re (z)}{dz}+(-\varepsilon +2jqz)\Re (z)=0  \label{eq:3}
\end{equation}%
where $\ell ,b,q$ and $\varepsilon $ are constants and $j$ takes integer and
half integer values. The function $\Re (z)$ is a polynomial of degree $2j.$
The differential equation can be obtained by introducing the following
linear and bilinear combinations of the generators of the $sl(2,R)$ Lie
algebra,%
\begin{equation}
\left[ J_{-}J_{0}+(\ell +j+1/2)J_{-}+qJ_{+}+bJ_{0}+(-\varepsilon +jb)\right]
\Re (z)=0  \label{eq:4}
\end{equation}%
which is quasi exactly solvable(QES)\cite{turb}. The differential
realizations of the generators of the algebra is given by\cite{turb}, 
\begin{equation}
J_{-}=\frac{d}{dz},\quad J_{0}=z\frac{d}{dz}-j,\quad J_{+}=-z^{2}\frac{d}{dz}%
+2jz.  \label{eq:5}
\end{equation}%
The function $\Re (z)$ forms a basis for $sl(2,R)$ Lie algebra. The solution
of the differential equation(\ref{eq:3}) \ which was determined in the paper%
\cite{koc} is in the following form%
\begin{equation}
\Re _{j}(z^{2})=\sum\limits_{m=0}^{2j}\frac{(2j)!(2\ell +1)!(\ell +m)!}{%
2m!(2j-m)!(2\ell +1+2m)!}P_{m}(\varepsilon )(-qz^{2})^{m}.  \label{eq:6}
\end{equation}%
Here the polynomial $P_{m}(\varepsilon )$ satisfies the recurrence relation%
\begin{equation}
(2j-m)qP_{m+1}(\varepsilon )-(\varepsilon -bm)P_{m}(\varepsilon )+m(\ell
+m+1/2)P_{m-1}(\varepsilon )=0  \label{eq:7}
\end{equation}%
with the initial condition $P_{0}(\varepsilon )=1.$ The polynomial $%
P_{m}(\varepsilon )$ vanishes for $m\geqslant 2j+1$ and the roots of $%
P_{2j+1}(\varepsilon )=0$ correspond to the $\varepsilon -$eigenvalues of
the algebraic Hamiltonian(\ref{eq:4}). It is well known that the
differential equation (\ref{eq:3}) can be transformed into the form of the
Schr\"{o}dinger equation and several quantum mechanical potentials can be
generated. In order to discuss all the potentials related to the
differential equation(\ref{eq:3}), in a unified manner we introduce a
variable $z=r(x)$ then the equation (\ref{eq:3}) takes the form: 
\begin{equation}
\frac{r}{r^{\prime 2}}\frac{d^{2}\Re (x)}{dx^{2}}+\frac{1}{r^{\prime }}\left[
\ell +3/2+r(b-qr)-\frac{rr^{\prime \prime }}{r^{\prime 2}}\right] \frac{d\Re
(x)}{dx}+(-\varepsilon +2jqr)\Re (x)=0  \label{eq:8}
\end{equation}%
Now lets turn our attention to the effective mass Schr\"{o}dinger equation(%
\ref{eq:2}). In this case both the Schr\"{o}dinger equation and the QES
differential equation (\ref{eq:8}) include first order differential terms.
One can easily transform the effective mass Schr\"{o}dinger equation into
the form of (\ref{eq:8}). It is convenient to express the eigenfunction $%
\psi (x)$ in the usual form%
\begin{equation}
\psi (x)=-\frac{2r}{r^{\prime 2}}m(x)e^{-\int W(x)dx}\Re (x).  \label{eq:9}
\end{equation}%
Substituting (\ref{eq:9}) into (\ref{eq:2}) and then comparing with (\ref%
{eq:8}) we obtain the following expression for the weight function $W(x)$%
\begin{equation}
W(x)=\frac{1}{4}\left( \frac{2m^{\prime }(x)}{m(x)}-\frac{6r^{\prime \prime }%
}{r^{\prime }}+\frac{(1-2\ell -2br+2qr^{2})r^{\prime }}{r}\right)
\label{eq:10}
\end{equation}%
and an implicit expression for the potential function, as follows%
\begin{align}
m(x)& \left[ V(x)-E\right] =  \notag \\
& \frac{(\beta +1/4+\alpha (\alpha +\beta +1))m^{\prime 2}(x)}{2m^{2}(x)}%
-\beta \frac{m^{\prime \prime }(x)}{4m(x)}+\frac{3}{8}\left( \frac{r^{\prime
\prime }}{r^{\prime }}\right) ^{2}-\frac{r^{\prime \prime \prime }}{%
4r^{\prime }}  \label{eq:11} \\
& \left( b^{2}-(2\ell +8j+5)q+\frac{4\varepsilon +b(2\ell +3)}{r}+\frac{\ell
(\ell +1)-3/4}{r^{2}}-2bqr+q^{2}r^{2}\right) \frac{r^{\prime 2}}{8},  \notag
\end{align}%
where $r^{i}$ is $i^{th}$ derivative of $r$ with respect to $x$.

At this point we first discuss the special form of the above potential. When
we choose $q=0$ the potential is exactly solvable. Under the conditions, $%
q=0 $ and $m(x)=$constant the potential leads to the Natanzon confluent
potentials\cite{chef}. To obtain the quantum mechanical potentials the
function $r(x)$ should satisfy the relation%
\begin{equation}
\sqrt{\lambda _{0}+\lambda _{1}/r(x)+\lambda _{2}/r^{2}(x)}\frac{dr}{dx}=-%
\sqrt{m(x)}.  \label{eq:12}
\end{equation}%
As for the special cases, $\lambda _{0}=\lambda _{2}=0,$ the potential
corresponds to the radial sextic oscillator potential; $\lambda _{1}=\lambda
_{2}=0$ to the QES Coulomb potential and $\lambda _{0}=\lambda _{1}=0,$ to
the Morse potential.

For the corresponding special cases we obtain the following mass dependent
potentials with some parameters. Let $\lambda _{0}=\lambda _{2}=0$ and $%
\lambda _{1}=1/4$ then $r(x)=-u^{2}=-\left[ \int \sqrt{m(x)}dx\right] ^{2}$
and the potential takes the form,%
\begin{eqnarray}
V(x) &=&\frac{\ell (\ell +1)}{2u^{2}}+\frac{1}{2}\left( b^{2}-(2\ell
+8j+5)q\right) u^{2}+bqu^{4}+\frac{1}{2}q^{2}u^{6}+  \notag \\
&&\frac{\left( \alpha (\alpha +\beta +1)+\beta +9/16\right) m^{\prime 2}(x)}{%
2m^{3}(x)}-\frac{(1+2\beta )m^{\prime \prime }(x)}{8m^{2}(x)}.  \label{eq:13}
\end{eqnarray}%
This is a family of radial sextic oscillator potential. We have checked that
the for choice of $q=0$ and $m(x)=\left( \frac{a+x^{2}}{1+x^{2}}\right) ^{2}$
the potential takes the same form as the potential given in the paper\cite%
{roy} and for $m(x)=cx^{2}$ the potential corresponds to the potential given
by Dutra\cite{dutra}. The eigenvalue of the Schr\"{o}dinger equation with
the potential given in (\ref{eq:7}) is given by%
\begin{equation}
E=\left( \ell +\frac{3}{2}\right) b+2\epsilon .  \label{eq:14}
\end{equation}%
The energy parameter $\varepsilon $ is obtained from the recurrence relation(%
\ref{eq:7}).

For the cases $\lambda _{1}=\lambda _{2}=0$ and $\lambda _{0}=1/4$ the
function $r(x)=-2u$ and the potential takes the form%
\begin{eqnarray}
V(x) &=&\frac{\ell (\ell +1)-3/4}{8u^{2}}-\frac{4\varepsilon +(2\ell +3)b}{4u%
}+2bqu+2q^{2}u^{2}+  \notag \\
&&\frac{\left( \alpha (\alpha +\beta +1)+\beta +9/16\right) m^{\prime 2}(x)}{%
2m^{3}(x)}-\frac{(1+2\beta )m^{\prime \prime }(x)}{8m^{2}(x)}.  \label{eq:15}
\end{eqnarray}%
This potential represents a family of QES Coulomb potentials. In order to
obtain the standard form of the potential one should redefine the
parameters. The eigenvalues of the potential is given by%
\begin{equation}
E=-\frac{1}{2}((2\ell +8j+5)q-b^{2})  \label{eq:16}
\end{equation}%
For the last example we choose $\lambda _{0}=\lambda _{1}=0$ and $\lambda
_{2}=1$ to obtain a family of QES Morse potentials. Then $r(x)=e^{-u}$ and
the potential takes the form%
\begin{align}
V(x)=& \frac{1}{2}\left( \varepsilon +(\ell /2+3/4)b\right) e^{-u}+\frac{1}{2%
}\left( b^{2}/4-(\ell /2+j+5/4)q\right) e^{-2u}-  \notag \\
& \frac{bq}{4}e^{-3u}+\frac{q^{2}}{8}e^{-4u}+  \label{eq:17} \\
& \frac{\left( \alpha (\alpha +\beta +1)+\beta +9/16\right) m^{\prime 2}(x)}{%
2m^{3}(x)}-\frac{(1+2\beta )m^{\prime \prime }(x)}{8m^{2}(x)}.  \notag
\end{align}%
The standard form of the Morse potential can be obtained by reordering the
parameters. The corresponding eigenvalue is given by%
\begin{equation}
E=-\frac{1}{8}\left( \ell (\ell +1)+1/4\right) .  \label{eq:18}
\end{equation}

We have constructed a class of QES potential for the generalized effective
mass Hamiltonian without any restriction in the parameters $\alpha $ and $%
\beta $. We have shown that one can obtain a family of potentials, related
to the sextic oscillator, QES Coulomb and QES Morse potentials. The method
discussed here can be used to obtain other class of potentials which can be
related to the hypergeometric Natanzon class potentials.

\end{document}